# DEPLOYING HEALTH MONITORING ECU TOWARDS ENHANCING THE PERFORMANCE OF IN-VEHICLE NETWORK

Geetishree Mishra[1], Rajeshwari Hegde[2], K S Gurumurthy[3]

[1]Dept of E&C Engg, BMS College of Engg, Bangalore, India
`geetishree@gmail.com`
[2]Dept of Telecommunication Engg, BMS College of Engg, Bangalore, India
`Rajeshwari.hegde@gmail.com`
[3]Dept of E&C Engg, UVCE, Bangalore, India
<drksgurumurthy@gmail.com>

## ABSTRACT

*Electronic Control Units (ECUs) are the fundamental electronic building blocks of any automotive system. They are multi-purpose, multi-chip and multicore computer systems where more functionality is delivered in software rather than hardware. ECUs are valuable assets for the vehicles as critical time bounded messages are communicated through. Looking into the safety criticality, already developed mission critical systems such as ABS, ESP etc, rely fully on electronic components leading to increasing requirements of more reliable and dependable electronic systems in vehicles. Hence it is inevitable to maintain and monitor the health of an ECU which will enable the ECUs to be followed, assessed and improved throughout their life-cycle starting from their inception into the vehicle. In this paper, we propose a Health monitoring ECU that enables the early trouble shooting and servicing of the vehicle prior to any catastrophic failure.*

## 1. INTRODUCTION

Today's automotive systems can be seen as sophisticated distributed systems where ECUs are networked to function efficiently in a coordinated manner. More functionalities related to safety and convenience are integrated into a vehicle system, there by increasing the number of ECUs which ranges from a dozen up to a hundred, supporting several subsystems, e.g., cruise control, powertrain management, and suspension control [1]. In the current scenario, high end cars are equipped with around 80 ECUs that send and receive more than 2500 signals using five different buses with proper gateways. These ECUs communicate and exchange reliable signals among them for the proper functioning of the vehicle [2]. The safety requirement of the function impacts directly the constraint of the signals exchanged between ECUs. Low and medium safety needs are already addressed in currently employed communication protocols with mechanisms like error detection. The role of fault tolerant communication is to address the higher communication safety needs. The fault tolerant communication increases the dependability of signal transmission between ECUs, sensors and actuators.

The design flow for automotive electronics includes a system specification phase, traditionally performed by car manufacturers, a subsystem implementation phase, traditionally performed by subsystem providers together with SW and HW component providers, and an integration phase, traditionally performed by the car manufacturer. Cost reduction requires integrating functionality from multiple suppliers onto a single ECU, while system integration requires interconnecting several ECU's, sensors, actuators using a network bus (e.g. CAN) and dedicated wiring. Often, the increasing number of ECU's is more a consequence of bad design practice (one ECU per sensor, local redundancy) rather than a real necessity. As a result, buses are needed to replace the otherwise large amount of wiring required to connect the ECU's altogether. In any case, both ECU SW integration and system integration are error prone, so far mostly manual procedures. Likewise, the proof of the system safety (e.g. for applications such as brake by wire) is often performed via fault injection in the lab. This process is not easily repeatable and therefore expensive [3]. Hence there arises a need for health monitoring ECU to take preventive measures. The paper is organized as follows. Section II deals with the need for Electronic Control Units in automotives. Section III deals with In-Vehicle Networks and protocols. Section IV emphasizes on the need of health monitoring of ECUs. Section V explains the proposed health monitoring ECU. The paper is concluded in section VI.

## 2. NEED FOR ECUS IN AUTOMOTIVES

The Importance of electronics is increasing day by day in modern automotives and the same has lead to proliferation of ECUs in them. It is an embedded system that controls one or more of the electrical subsystems in a vehicle. The comfort, safety and performance of a modern car depends largely on these ECUs. The development of a new ECU is a joint effort of the car manufacturers and the suppliers. The car manufacturer writes a specification of the desired functionality for which the software and hardware is delivered by the supplier [4]. The development of ECUs is part of the development of a whole car. The most complex ECUs operates the powertrain. Simpler ones operate functions such as power windows, power seat, mirror adjustment system, etc. But even these ECUs need to be networked so that those specific features can be exploited both from the view of power management and such other critical co-ordinations or for enhancing the utility by way of personalization, etc. This trend of increasing automotive electronic content is the direct result of many new features that will greatly increase both safety and comfort but that will require more sophisticated ECUs with a large embedded software component. The safety features include X by-wire, automatic lane-following, drowsy driver detection, intelligent cruise control and airbag systems that can adjust deployment based upon passenger weight and the specific nature of an accident. Improving fuel efficiency is an important goal: hybrid and fuel-cell electric drive place high demands on software. Telematics and in-car entertainment will further increase the electronic content of cars, requiring the combination of such technologies as wireless connectivity, global positioning, digital radio and Internet access, all with hands-free voice activation whenever possible. [5]. For example, Antilock Brake Systems (ABS), which were introduced in 1978 and became widespread in the '80s, marked the first of the car overrides- driver technologies and ABS laid the foundation for all the systems that have followed. Essentially, ABS works by using sensors to keep a central electronic control unit apprised of the rotating speed of each wheel. The processor regulates hydraulic pressure to apply maximum braking force without causing a skid. Building on that data and hardware, engineers increased the capabilities of the computer controls to create traction control in the '90s and, soon after, Electronic Stability Control (ESC). It keeps the vehicle from either over steering or under steering and keeps the vehicle stable when pressure is applied to individual brakes. When the driver finds him/herself in a critical situation and the vehicle threatens to skid, ESC intervenes autonomously. With ESC the driver can maintain control of the vehicle in critical situations so that a crash can be prevented in many cases. [6].

## 3. IN-VEHICLE NETWORK AND PROTOCOLS

In the early days of automotive electronics, each new function was implemented as a stand-alone ECU, which is a sub system composed of a microcontroller and a set of sensors and actuators. This approach quickly proved to be insufficient with the need for functions to be distributed over several ECUs and the need for information exchanges among functions. For example, the vehicle speed estimated by the engine controller or by wheel rotation sensors has to be known in order to adapt the steering effort, to control the suspension, or simply to choose the right wiping speed [7]. Until the beginning of the 1990s, data was exchanged through point-to-point links between ECUs. However this strategy, which required an amount of communication channels of the order of the number of ECUs was unable to cope with this increase due to the problems of weight, cost, complexity, and reliability induced by the wires and the connectors. These issues motivated the use of networks where the communications are multiplexed over a shared medium, which consequently required defining rules, protocols for managing communications and, in particular, for granting bus access [8]. In order to handle the increasing complexity of the in-car network system engineering and architecture was introduced, leading to so-called domain oriented bus topologies. The electronic systems of a car are divided into domains like body electronics or vehicle motion management. Within such a domain a so-called domain controller takes care of managing the tasks (e.g. reading sensor outputs, sending messages to actuators) within the domain. This controller connects intelligent sensors and actuators via local bus systems and is by itself connected via a central gateway to other domain controllers [9]. The functions within a vehicle are divided into systems and sub-systems to provide for passenger entertainment, comfort, and safety, as well as to improve vehicle performance and enhance powertrain control. These systems must communicate with one another over a complex heterogeneous In-Vehicle Network (IVN). Each network typically contains multiple communication protocols including the industry standard Controller Area Network (CAN), Local Interconnect Network (LIN), FlexRay and MOST(Media Oriented System Transport) and Byteflight[10]. F. Baronti et al. discussed the hardware implementation of high speed and fault-tolerant communication systems for in-vehicle networking [11]. The demand for drive-by-wire, telematics, entertainment, multimedia, pre-crash warning, remote diagnostic and software update, etc. will significantly increase the complexity of the future in-vehicle communication networks. Figure 1 shows the schematic of the In-Vehicle Network used in current vehicles [12].

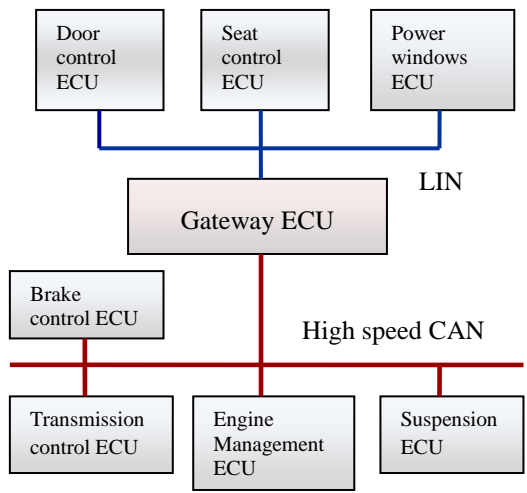

Figure1. Schematic of the current In-Vehicle Network.

CAN is one of the most popular in-vehicle network. It simplifies the cable in vehicle, so the weight of the vehicle is reduced and easy to maintain and repair. Many ECUs are connected to CAN network and communicate with each other [13]. A low-speed master-slave time triggered protocol meant to connect on-off type loads to higher speed networks. Typical loads include door locks, sun roofs, rain sensors, and powered mirrors. A LIN network is used as a low cost alternative if the full functionality of the CAN protocol is not required [14].

FlexRay is a fault-tolerant high-speed communication protocol targeted for safety critical applications such as steer-by-wire, brake-by-wire etc. It has a maximum data rate of 10 Mbps. Along with enabling safety-related applications, a FlexRay network is well suited as a communication backbone connecting heterogeneous networks together [15].
MOST is multimedia network in vehicle, and was proposed by BMW, Harman/Becker, DaimlerChrysler, OASIS Silicon System, etc [16]. It was developed to provide in-vehicle multimedia and infotainment systems with communication during the transmission of audio, video, data and control information. The MOST cooperation, a consortium of car makers, system architects and key component suppliers, started to develop a multimedia network in 1998, and now MOST is the de-facto standard for such applications[17]. Byteflight is a high performance communication network developed for safety related applications in passenger vehicles, which features the advantages of both the synchronous and asynchronous transmission schemes [18]. The protocol unites the advantages of the synchronous and asynchronous methods. It also guarantees deterministic latencies for a specific number of high priority messages and flexible use of transmission bandwidth for low-priority messages [19].

## 4. NEED FOR HEALTH MONITORING OF ECUS

Over the years, the exponential increase of automotive electronics led to reduce the number of casualties from accidents even with the increase of traffic density. The emission of cars has been reduced tremendously with the help of new electronic control units, new sensor technologies and electronically controlled actuators. At the same time fuel consumption also decreased by vehicles. These are the driving forces for employing huge number of new electronic functions which are the basis of all these ever required improvements in automotives. New high end cars have upto 80 ECUs assembled and networked to work as a unified distributed system. ECUs have more complex and sophisticated embedded software both for monolithic and distributed system support. Managing the increasing complexity due to the increasing number of ECUs and the embedded software, became the most challenging job for the automotive manufacturers. ECUs being in a network exchange critical information among them to work in a collaborative way to meet the time constraints. So more specifically we can say that as ECUs have become indispensable for any automotive electronic system, they should be available and running healthily as the vehicle runs. So here comes the need of continuous health monitoring of ECUs with the start of the vehicle. In this paper we propose a health monitoring ECU which can be implemented to fulfill the requirements.

## 5. PROPOSED SYSTEM FOR HEALTH MONITORING OF ECU

In critical real-time system, in which timeliness that is the ability of a system to meet time constraints such as deadlines is significant, avoidance of failure is the only option. Since the ECUs in a vehicle are real time systems and function continuously when a vehicle is driven, they have to be monitored for the correctness of the system to make them available always. This necessitates the health monitoring ECU which monitors the health conditions of ECUs in a network in a cyclic manner. In this paper, we propose a Health monitoring ECU which will generate event message commands in a cyclic manner for each ECU and receives an acknowledgement periodically from each within the allocated time. As long as the health monitoring ECU receives positive acknowledgement, it infers that the ECUs in a network function as desired. Once it receives negative acknowledgement with the error message ID, it starts the necessary corrective action if it is a minor error that can be dealt with. It also sends a warning message to the on-board display unit. In worst case scenario, if it does not receive any acknowledgement from the other end, the error message about the faulty ECU is intimated to the driver as a notification and as well as to a nearby service station. This enables the early trouble shooting and servicing of the vehicle prior to any catastrophic failure. The figure shows the flow chart of the proposed system.

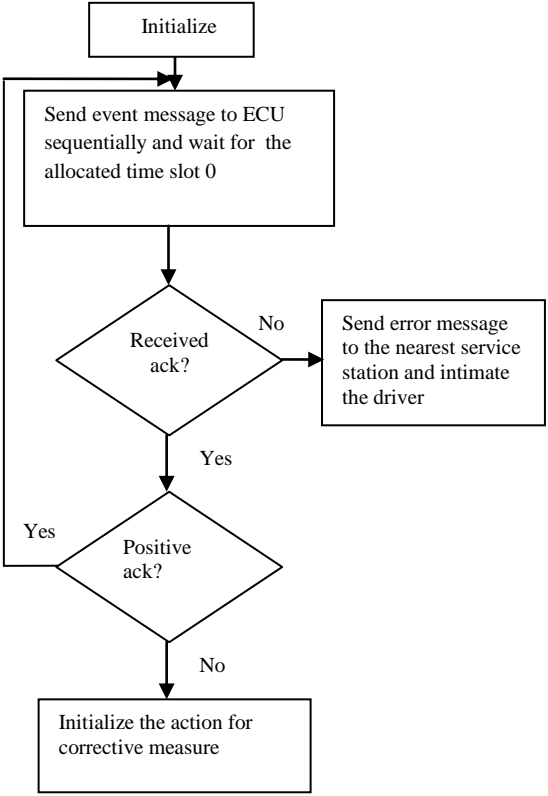

Figure 2. Work flow diagram of Health Monitoring system

The hardware and software requirements for the working of health monitoring ECU are medium computational capability controller with minimum memory because the main self testing software will be running in every ECU and the health monitoring ECU has to execute a small code to communicate with each of these by sending event command messages with corresponding message ID and by receiving acknowledgement from the same. It is also required to communicate about the error to the external service station. So it needs to have a wireless enabled on board device. Since this health monitoring ECU should also display the error messages to the driver, it has to have a display unit as shown in the fig 3.

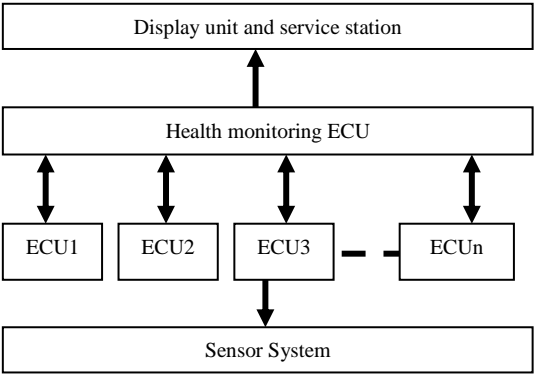

Figure3. Block Diagram Of Health Monitoring System

The health monitoring ECU being in a network validates the operating conditions of all the networked ECUs and monitors vehicle performance every time it is driven, identifying the performance related problems immediately and providing technicians with information to help them quickly and accurately diagnose and repair malfunction [20]. It also reports the position and vehicle information at a regular interval to the nearby service station as shown in the figure 4.

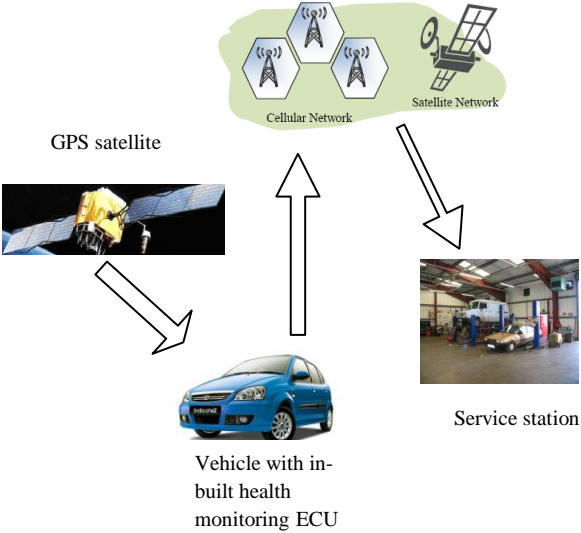

Figure 4. Deployment of Health monitoring ECU

# 6. CONCLUSION

In this paper, the need for health monitoring of ECUs is discussed. The functions of proactive health monitoring ECU in the multi ECU configuration is highlighted. The proposed health monitoring ECU can value add for preventing the catastrophic failure of the vehicle. This can reduce the maintenance cost of the vehicle and ensures that the vehicle doesn't fail instantaneously and will optimize the real time performance of a vehicle.

## Authors

**Rajeshwari Hegde** completed her Bachelor of Engineering in Electronics and Communication Engineering from National Institute of Engineering, Mysore and Master of Engineering in Electronics from BMS College of Engineering, Bangalore  She has completed her research in the field of automotive embedded system  at Bangalore University, India,  under the guidance of Dr K S Gurumurthy. Her major field of study includes embedded system, signal processing and wireless communication. She is currently working as Associate Professor in the Department of Telecommunication Engineering, BMS College of Engineering, Bangalore, India. She has published 42 research papers in international conferences, national conferences and reputed journals. 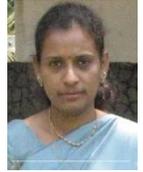

**Geetishree Mishra** completed her AMIE in  Electronics and Communication Engineering from The Institution of Engineers, India, Master of Technology in Digital Communication from BMS College of Engineering, Bangalore, India.  She has  seven  years of industry experience from Radiant Telesystems Limited  Bhubaneshwar, Orissa and teaching experience of three years. Currently she is working as Assistant Professor in the department of Electronics and Communication Engineering, BMS College of Engineering, Bangalore. She has published 5 research papers in international conferences and reputed journals. Her research interests include Automotive Embedded Systems, Signal Processing, In-Vehicle networks. 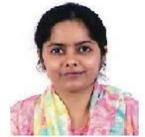

**Dr K S Gurumurthy** completed his Bachelor of Engineering in Electronics and Communication , from Mysore University and Master of Engineering in Solid State Electronics from U.O.R., Roorkee (now IIT, Roorkee),  and PhD from Indian Institute of Science, Bangalore, India. His major field of study includes VLSI and  microelectronics. 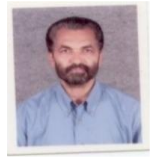

  He has two years of industrial experience and more than 30 years of research and academic experience. He is currently working as Professor and Head, Department of Electronics and Communication Engineering, University Visvesvarayya College of Engineering, Bangalore, India. He has given more than 50  guest lectures on different topics of VLSI    VLSI DESI at various colleges, WIPRO,CRL and at  IETE. He has published more than 85 research papers in various international journals and conferences.   His research interests are VLSI Design, Embedded System and Communication. Dr Gurumurthy is the professional member of IEEE and  ISTE.  He is the recipient of KHOSLA AWARD for the BEST Technical Paper presented at UOR, Roorkee, India.